\documentclass{aastex}
\usepackage{emulateapj5}
\usepackage{epsfig}
\usepackage{graphicx}

\newcommand{\bq}{\begin{equation}}
\newcommand{\eq}{\end{equation}}

\shorttitle{Structure and evolution of starburst galaxies}
\shortauthors{Mobasher {\it et al}}

\begin{document}

\title{ Structure  and Evolution of Starburst and Normal Galaxies}
\author{Bahram Mobasher\altaffilmark{1,2}, Shardha Jogee\altaffilmark{1}, Tomas Dahlen\altaffilmark{1}, Duilia De Mello\altaffilmark{3}, Ray A. Lucas\altaffilmark{1}, Christopher J. Conselice\altaffilmark{4}, Norman A. Grogin\altaffilmark{5}, Mario Livio\altaffilmark{1}}
\altaffiltext{1}{Space Telescope Science Institute, 
3700 San Martin Drive, Baltimore MD 21218, USA; b.mobasher@stsci.edu}
\altaffiltext{2}{Also affiliated to the Space Sciences 
Department of the European Space Agency}
\altaffiltext{3}{Goddard Space Flight Center, Greenbelt MD 20771}
\altaffiltext{4}{California Institute of Technology, Mail Code 105-24, Pasadena, CA 91125}
\altaffiltext{5}{Physics and Astronomy Department, Johns Hopkins University, Baltimore, MD 21218}
\begin{abstract}
A comparative study of the rest-frame morphology and structural properties
of optically selected starburst galaxies at redshift $z \la 1$ is 
carried out using multi-waveband ($BViz$) high resolution 
images taken by the Advanced Camera for Surveys (ACS) as  
part of the Great Observatories Origins Deep Survey (GOODS). 
We classify galaxies into starburst, early and late types
by comparing their observed spectral energy distributions (SEDs) 
with local templates. We find that 
early-type systems have significantly higher rest-frame $B$-band  
concentration indices  and  AGN fraction  ($>$25 \%)  
than late-type spirals and optically-selected starbursts. 
These results are  consistent with the scenario that early-epoch
($z \gg 1$) gas-rich dissipative processes (e.g., major mergers) 
have  played an important role in developing  large central 
concentrations in early-type E/Sa galaxies and that 
a concurrent growth of central black holes and bulges 
occur in some of these early merger events.   
The  lower  AGN fraction and concentration indices in the  majority of 
the optically-selected starbursts at $z \la 1$ 
suggest  that  either the starbursts and early types  
are different in  nature (being respectively disk and bulge dominated), 
or/and  are in different evolutionary phases  such that some of the 
starbursts in major mergers evolve into early-types as the dynamical
phase of the merger evolves and the spectral  signature of the starburst 
fades out.  The starbursts have, on average, larger asymmetries than 
our control sample of normal galaxies in both rest-frame $B$ 
and $R$-bands, suggesting 
that a significant fraction of the starburst activity is tidally triggered.  

\end{abstract}

\keywords{galaxies: evolution --- galaxies: 
starburst: galaxies}

\section{INTRODUCTION}

A coherent picture of the formation and evolution of galaxies 
can only emerge through an understanding of starbursts  over a 
wide range of cosmic lookback time.
Starbursts  drive  galactic superwinds (e.g., Lehnert \& Heckman 1996)
which play a key role in the metal  enrichment of 
the intergalactic medium and the establishment of the strong 
mass-metallicity relation in ellipticals  and bulges (Lynden-Bell 1992).
Moreover, the evolution of starbursts and  active galactic nuclei (AGN) 
may be  
intimately linked  through a common central gas supply and  
evolutionary  effects (e.g., Norman \& Scoville 1988; Kauffmann 
\& Haehnelt 2000). 

To date, detailed studies of star-formation have primarily targeted 
galaxies at high redshifts ($z\sim 2.5-5$) using
optical  (e.g., Steidel et al. 1999) and submillimeter 
(e.g., Hughes et al. 1998; Barger et al. 2001) emission. Conversely, 
in the local Universe ($z < 1$), studies of starbursts have mainly
focused on dusty ultra-luminous infrared galaxies (ULIRGs) which show
signs of strong interaction or double nuclei (Joseph \& Wright 1985;  
Armus, Heckman \& Miley 1987; Scoville et al 2000)
and large gas densities (Sanders, Scoville, Soifer 1991). 
These are likely counterparts of high 
redshift sub-mm sources (Chapman et al 2003). 

This paper focuses on a population of starburst galaxies which is 
different from those studied before. The sample of starbursts
studied here is optically selected, complete  to  $R_{AB}=24$ mag, 
and extends to $z\sim 1$. Therefore, the sample preferentially 
consists of relatively dust-free blue starbursts which  differ in 
nature or in evolutionary stage from those selected at infrared wavelengths. 
Using multi-waveband high resolution 
images from the Advanced Camera for Surveys (ACS), obtained as a part of
the Great Observatories Origins Deep Survey (GOODS), we compare the  
optically-selected starbursts at $z \la 1$ 
to a control sample consisting of normal late and early 
type systems,  and discuss their role  in the evolution of these galaxies.

\section{Observations and Sample Selection}

\subsection{Observations}

The GOODS project consists primarily of  space-borne ($HST$/ACS) 
and ground-based multi-waveband imaging of two fields, the HDF-N and 
CDF-S. In this study we consider only the Southern GOODS field (CDF-S)
where most of the initial data have been obtained and processed. 

The GOODS-ACS observations are performed in four bands ($BViz$), 
covering an area of $10 \arcmin \times 16 \arcmin$.
The ground-based imaging observations cover the entire CDF-S area
in $U$'$UBVRIJHK$ to the depths of $R_{\rm AB}$=25 mag 
and $K_{\rm s_{\rm AB}}$=23 mag. 
A more detailed discussion of the GOODS observations is given in 
Giavalisco et al (2003).

\subsection{Photometric Redshifts and Spectral types}

The ground-based CDF-S data provide photometry in 14 passbands consisting 
of WFI ($U$'$UBVRI$), FORS ($RI$), SOFI, and ISAAC ($JHK$). The observations
cover the entire field except for the FORS and ISAAC observations which 
are deeper and  only cover a sub-area of the field. Magnitudes measured
over an aperture with a $3 \arcsec$ diameter were used to construct 
the observed SEDs for all galaxies in the $R$-band selected survey. 
Photometric redshifts and spectral types were estimated for individual 
galaxies by fitting the observed and template SEDs using the Bayesian 
photometric redshift technique (Benitez 2000). The template SEDs 
consist of E, Sbc, Scd, and Im (Coleman, Wu \& Weedman 1980), and 
starburst (Kinney et al 1996).  A sub-sample of galaxies in the 
CDF-S have spectroscopic redshifts from the K20 survey 
(Cimatti et al 2002). Comparison between the estimated  
photometric and spectroscopic redshifts for these galaxies gives 
${z_{phot}-z_{spec}\over (1 +z_{spec})}\sim 0.10$, which is taken as the error
in photometric redshifts here. Additionally, the derived spectral types of
the galaxies were found to be consistent with visually classified 
elliptical and spiral galaxies on the ACS images. 
A detailed study of the photometric redshift measurements for the GOODS field
will be presented in Mobasher et al (2003).

\subsection{Sample Selection}

The starburst sample in this paper includes galaxies from the $R$-band 
selected ground-based survey which satisfy the following criteria: 
(a) They have starburst spectral types and  $R_{\rm AB} < 24$ mag.
The magnitude limit ensures that photometric redshifts and spectral 
types are measured more accurately (${\Delta(z)\over (1+z_{spec})} < 0.1$).
(b) They have redshifts in the range $ 0.2 < z < 1.3$.  
Lower redshifts are excluded to minimize  relative uncertainties in 
photometric redshifts and resolution dependent effects (see $\S$ 3). 
(c) They lie in the CDF-S area covered by the ACS.
A control sample  was also constructed (to the same 
magnitude limit) consisting of normal galaxies (i.e. non-starbursts), 
with early (E/Sa)  and late  (Sb--Sd) spectral types.
Im galaxies are not included.

Given the selection criteria,  our sample of starburst galaxies 
is biased towards  relatively dust-free blue starbursts and 
against dusty star-forming galaxies. 
This study  thus complements previous investigations of 
dusty starbursts  and ULIRGs which were  selected based on  
their infrared properties (e.g., Scoville et al 2000).
It is to be noted that the SED fitting method used here 
does not fully guarantee that the selected objects are indeed 
starbursts. For example, a very metal poor galaxy with only modest 
amount of star formation could have a very blue
color and hence, be classified as a starburst here. However, the number
of such galaxies is expected to be small with the sample mostly 
dominated by galaxies experiencing bursts of star formation.

Table 1 presents the number of starburst and control galaxies 
in their observed frames, so that in the redshift intervals indicated, 
they correspond to the rest-frame $B$-band. The median absolute magnitudes
for the three different types of galaxies in the present sample 
($R_{\rm AB} < 24$) are $M_B= -19.6$ (starburst), $-19.9$ (Sb--Sd) 
and $-20.6$ (E/Sa). 
The median $U-V$ ($V-K$) colors for the starburst, late and early-type
galaxies here are respectively estimated as $-0.70$ (1.80), $-0.20$ (2.40)
and 0.35 (2.8). 
In order to check the sensitivity of our results 
to the adopted magnitude limit, the analysis in this letter ($\S$ 3) was  
repeated to $R_{\rm AB} <$ 25 mag, with no difference found.

\section{ Structure and Morphology }

We quantify the concentration and asymmetry  
of the sample galaxies using the CAS concentration ($C$) 
(Bershady et al. 2000) 
and asymmetry ($A$) indices (Conselice et al. 2000). 
The concentration  index is proportional to  the logarithm of the 
ratio of the 80 \% to 20 \% curve of growth radii

\begin{equation}
C =  5 \times  log (r_{80 \%}/r_{20 \%}).
\end{equation}

The asymmetry index is obtained by rotating a galaxy image by  $\pi$,
subtracting it from its pre-rotated image, summing the intensities of
the absolute value  residuals, and normalizing the sum to the 
original galaxy flux. The concentration 
and asymmetry parameters are measured for the starburst sample in the
rest-frame $B$-band ($C_ {\rm B}$, $A_ {\rm B}$) to $z\sim 1$, 
using different observed 
ACS bands in the appropriate redshift ranges, as shown  in Table 1. 
Corresponding quantities were also derived in 
the rest-frame $R$-band ($C_ {\rm R}$, $A_ {\rm R}$). The
rest-frame $B$ and $R$ bands are respectively sensitive to young stars and
the underlying  older stellar population.  

In order to minimize  resolution dependent effects and relative 
uncertainties in photometric redshifts,  we constrain
the redshift range to $z \sim $ 0.2--1.3, where 
the change in spatial resolution ($ 0.05 \arcsec $) is only 
a factor $\sim$ 2, ranging from  160 pc to 400 pc, assuming
a flat cosmology with $\Omega_M = 1 - \Omega_{\Lambda} =0.3 $ 
and a Hubble constant $H_0$ = 70 km s$^{-1}$  Mpc$^{-1}$. 
The structural  comparisons in $\S$ 3.1--3.2  probe galaxy evolution 
over a time period where the age of the Universe varied from 6 to 11 
Gyr. It is important to note that the starburst and  
control samples are defined purely based on SED fits  to multi-band 
ground-based  images, without using  any morphological information. Thus, 
the derived structural parameters provide additional 
\it independent probes  \rm  of the nature and evolutionary stages 
of these systems.

\subsection{Concentration and AGN fraction}

Figure 1  shows the distribution of rest-frame $B$  concentration 
indices ($C_ {\rm B}$)  for the starburst galaxies and the 
control sample, divided into two redshift bins.  
The  distribution of $C_ {\rm B}$ in early-type systems is  
significantly skewed towards higher values compared to starburst 
galaxies. Results from both the $T$-test and the Kolmogorov-Smirnov test, 
listed in Table 2, confirm that the distribution of concentration indices 
for starburst and early-type galaxies are statistically different, 
both in the mean and the overall distribution. About 72 \% of the 
early-type galaxies have $C_ {\rm B}$ $>$ 3.0 compared to 12 
\% of the starburst  and 18 \% of the late-type galaxies (Table 2).

The data suggest that by $z\sim$ 1, early-type systems  have 
already developed large central light concentrations, which demarcate 
them from optically selected starbursts and late-type  galaxies 
(Figs. 1a and 1b). 
These large concentrations  are likely  associated with bulges and 
spheroidal components. This was demonstrated by simulation of galaxies
with de Vaucouleurs (r$^{1/4}$) law, which were found to have large
$C_ {\rm B}$ ($>$ 3.4) values (Jogee et al. 2003, in prep.).  
Furthermore, comparison of  our samples with the CDF-S X-ray catalog
(Alexander et al 2003) shows that  a much larger fraction ($>$ 25 
\%) of the early-type systems host AGNs  compared to 
 only 2\% of the starbursts\footnote{Assuming
$ 1.5 arcsec$  as the maximum distance between the ACS $B$-band 
coordinates and the X-ray identified sources (Koekemoer et al. 2003). 
Extending the search to $6.0 \arcsec$ increases the number of 
starbursts to 7\%.}. 
The larger  AGN fraction and high $C_ {\rm B}$ we find in early-types 
are in qualitative agreement with the high concentrations in 
AGN hosts reported by other studies (e.g., Grogin at al. 2003).
One general scenario consistent with our 
results  is that the early-type systems with large  $C_ {\rm B}$ 
developed their central structures  (bulges and spheroids)  in much earlier 
($z \gg $1)  violent  gas-rich dissipative episodes such as major mergers 
(e.g., Naab \& Burkert 2001).  A  concurrent growth of bulges/spheroids 
and massive black holes may occur in some of these  mergers (Kauffmann 
\& Haehnelt 2000), consistent with the large AGN fraction in the early 
types.

What is the nature of the  optically selected starbursts at $z \la 1$ 
and what will they evolve into? Most  of these galaxies 
have concentrations in the range $C_ {\rm B}$=3.3--4.5,  which are well 
below those shown by the majority ($>$ 50 \%) of 
early-type galaxies, but comparable to those of late-type 
spirals in the same redshift range. 
Low values in $C_ {\rm B}$ can result from   disk-like systems as 
revealed by fits to artificial galaxies  with exponential disks 
(Jogee et al. 2003, in prep.).   However, 
$C_ {\rm B}$ in starbursts is more strongly influenced 
by very young stars  than in early-type systems so that 
low $C_ {\rm B}$  may also result from localized, off-centered 
star-forming regions which often exist in early stages of tidal 
interactions or mergers.  
Visual inspection of the starbursts confirms that they include 
disk-like systems as well as a large number of  morphologically 
disturbed  objects which show tails, double nuclei, and nearby 
companions with a range of luminosity ratios (1:10 to 1:1). 
In summary, the  lower  $C_ {\rm B}$ and AGN fraction in the  
majority of  optically-selected starbursts, compared to early-type 
systems, suggest two possibilities.  Either the starbursts 
and early types  are entirely different in  nature (being respectively 
disk and bulge dominated), or/and  are in different evolutionary 
phases  (that some of the starbursts in major mergers evolve 
into early-types as violent relaxation sets in  and starburst spectral 
signature fades out).

\subsection{Asymmetry}

Figure 2a  shows the distributions of rest-frame asymmetry ($A_{\rm B}$)
indices in the rest-frame $B$-band for the starburst and control sample. 
A larger fraction (50 \%) of  the starburst galaxies 
have a high $A_ {\rm B}$ $>$0.3,  compared to early-types (13 \%) 
and late-types (27 \%) (Table 2). 
It is likely that large values of $A_ {\rm B}$ are
driven  by  irregular patterns of star formation which may either  be  \it  
externally \rm  triggered  in tidally interacting/merging systems 
or \it spontaneously \rm triggered in isolated galaxies. 
Asymmetries induced by dust lanes, caused
by high inclinations, are not significant as most objects 
in our sample are moderately  inclined, based on their 
estimated ellipticities.
Fig. 2b  shows the corresponding  rest-frame  $A_ {\rm R}$
index, derived out to $z \sim 0.58$ (Table 1). 
$A_ {\rm R}$ is  less sensitive than the rest-frame $B$-band 
to massive young stars. Thus, tidally induced features in mergers 
(e.g., Jogee, Kenney, \& Smith 1999), can lead to large 
values of $A_ {\rm R}$. The presence of both large  
$A_ {\rm B}$ and  $A_ {\rm R}$ in a large fraction of the starbursts 
suggests that a significant part of their SF is  tidally triggered. 
High asymmetry values have also been previously noted in a local 
sample of infrared luminous starburst galaxies (Conselice et al 2000).  
Visual inspection  of our images confirms this claim as the majority  
of the systems  with large $A_ {\rm B}$ ($>$0.4) and $A_ {\rm R}$ 
($>$0.35) indeed  appear to be optically disturbed or/and to have 
nearby companions. 

\section{Summary}

From a  comparative study of optically-selected starburst galaxies 
out to $z \la 1$  and a control sample of normal (early-type E/Sa and 
late-type) galaxies, we  find:  

\noindent 
\bf (1) \rm The  early-type systems have  rest-frame $B$ 
concentration indices  and  AGN fraction  ($>$25 \%) which 
are  significantly higher than those of late-type spirals and  
optically-selected starbursts at $z \la 1$ . The results are generally 
consistent with the idea that early-epoch ($z\gg 1$) gas-rich 
dissipative processes (e.g., major mergers)  have likely  played an 
important role in developing  large central concentrations 
in early-type E/Sa galaxies. The larger AGN fraction 
qualitatively supports scenarios leading to a  concurrent growth 
of central black holes and bulges in some early merger events.

\noindent 
\bf (2) \rm 
The  lower  $C_ {\rm B}$ and AGN fraction  in the  majority of 
$z \la 1$ optically-selected starbursts, compared  to early-type 
systems,  suggest  that  either the starbursts and early types  
are different in  nature (being respectively disk and bulge dominated), 
or/and  are in different evolutionary phases  
(such that some of the starbursts in major mergers evolve into early-types).

\noindent 
\bf (3) \rm 
The $z\la 1$ optically-selected starbursts, on average, have larger 
asymmetry indices than normal galaxies, both in rest-frame $B$ and $R$-band.
This suggests that a significant fraction of the 
starbursts may be tidally triggered.

\acknowledgments 
Support for this work was
provided by NASA through grant GO09583.01-96A from the Space Telescope
Science Institute, which is operated by the Association of
Universities for Research in Astronomy, under NASA contract
NAS5-26555.



\clearpage
\begin{deluxetable}{cccccc}
\tabletypesize{\scriptsize}
\tablewidth{0pt}
\tablecaption{Number of starburst and non-starburst (control) galaxies
 from the ACS BV{\it iz} images \label{tbl-1} }
\tablecolumns{8}
\tablehead{
\colhead {Redshift range} &
\colhead {Filter} &
\colhead{Starbursts} &
\multicolumn{3}{c}{Non-Starbursts} \\
\colhead{} &
\colhead{} &
\colhead{} &
\colhead {All} &
\colhead {Early} &
\colhead {Late} \\
}
\startdata 
\multicolumn{6}{c}{ Rest-Frame $B$-band} \\
  $ 0.24<z<0.56 $ & F606W  &     54      &   451 &  100 &  351 \\
  $ 0.61<z<0.93 $ & F775W  &     102     &   377 &  122 &  255 \\
  $ 0.94<z<1.31 $ & F850LP &     1       &   149 &   21 &  128 \\
                 &        &     157      &   977 &  243 &  734 \\
\multicolumn{6}{c}{ Rest-Frame $R$-band} \\
  $ 0.20<z<0.32 $  & F775W  &    6       &  164 &   25  & 139 \\
  $ 0.33<z<0.58 $  & F850LP &    50      &  391 &   89 &  302 \\
                   &        &    56      &  555 &  114 &  441 \\
\enddata 
\end{deluxetable}

\begin{deluxetable}{ccccc}
\tabletypesize{\scriptsize}
\tablewidth{0pt}
\tablecaption{Comparison of Structural Parameters  for Starbursts and  Non-Starbursts.\label{tbl-2}}
\tablecolumns{8}
\tablehead{
\colhead{} &
\colhead{Starbursts} &
\multicolumn{3}{c}{Non-Starbursts} \\
\colhead {  } &
\colhead {  } &
\colhead {All  } &
\colhead {Early} &
\colhead {Late} \\
}
\startdata 
\multicolumn{5}{c}{ Rest-Frame $B$-band  } \\
Mean $C_ {\rm B}$     & 2.53 &  2.82   &  3.25   &   2.65 \\
Student T-test$^{1}$  &      &  3e-9   & 6e-28   &   4e-4 \\
K-S Test$^{2}$        &      & 0.10    & 7e-7    &   0.53  \\
$C_ {\rm B} > $ 3.00  & 12\% &  31\%   &  72\%   &   18\% \\
       &  & & & \\
Mean   $A_ {\rm B}$   & 0.33 &  0.26   &   0.23  & 0.27 \\
Student T test$^{1}$  &      &  9e-4   &  7e-11  & 3e-2 \\
K-S Test${^2}$        &      &  1e-6   &  1e-10  & 3e-4    \\
$A_ {\rm B}> $ 0.30   & 50\% &  23\%   &   13\%  & 27\%  \\

\multicolumn{5}{c}{ Rest-Frame $R$-band  } \\
Mean $C_ {\rm R}$     & 2.59 &  2.81   &  3.26  &   2.72 \\
Student T-test$^{1}$  &      &  3e-9 & 6e-28  &  4e-4 \\  
K-S Test$^{2}$        &      &  0.43 & 9e-5   & 0.77 \\
$C_ {\rm R} > $ 3.00  & 8\%  &  29\%   &  71\%  &  18\% \\
       &  & & & \\
Mean   $A_ {\rm R}$   & 0.31 &  0.24  &   0.22  &    0.24 \\
Student T test$^{1}$  &      & 8e-4   &   5e-4  &    3e-3 \\
K-S Test$^{2}$        &      & 0.24   &   5e-4   &   2e-2  \\
$A_ {\rm R} > $ 0.30  & 37\% &  15\%  &   15\%  &    15\%  \\

\enddata 
\tablecomments { 
(1) The significance S of the  Student's T-test 
   when comparing  the starburst  sample to each of the control
   populations  in   columns 2 3 and 4.  A  small value (e.g.,$<$0.05) 
   of S indicates  that the 2 populations have significantly different means; 
(2) As in (1) but showing the significance  in Kolmogorov-Smirnov 
   ($K-S$) test. This is  the significance level for the null hypothesis 
   that the two  data sets are drawn from the same distribution;
}
\end{deluxetable}

\clearpage

\begin{figure} 
\centerline{
\psfig{figure=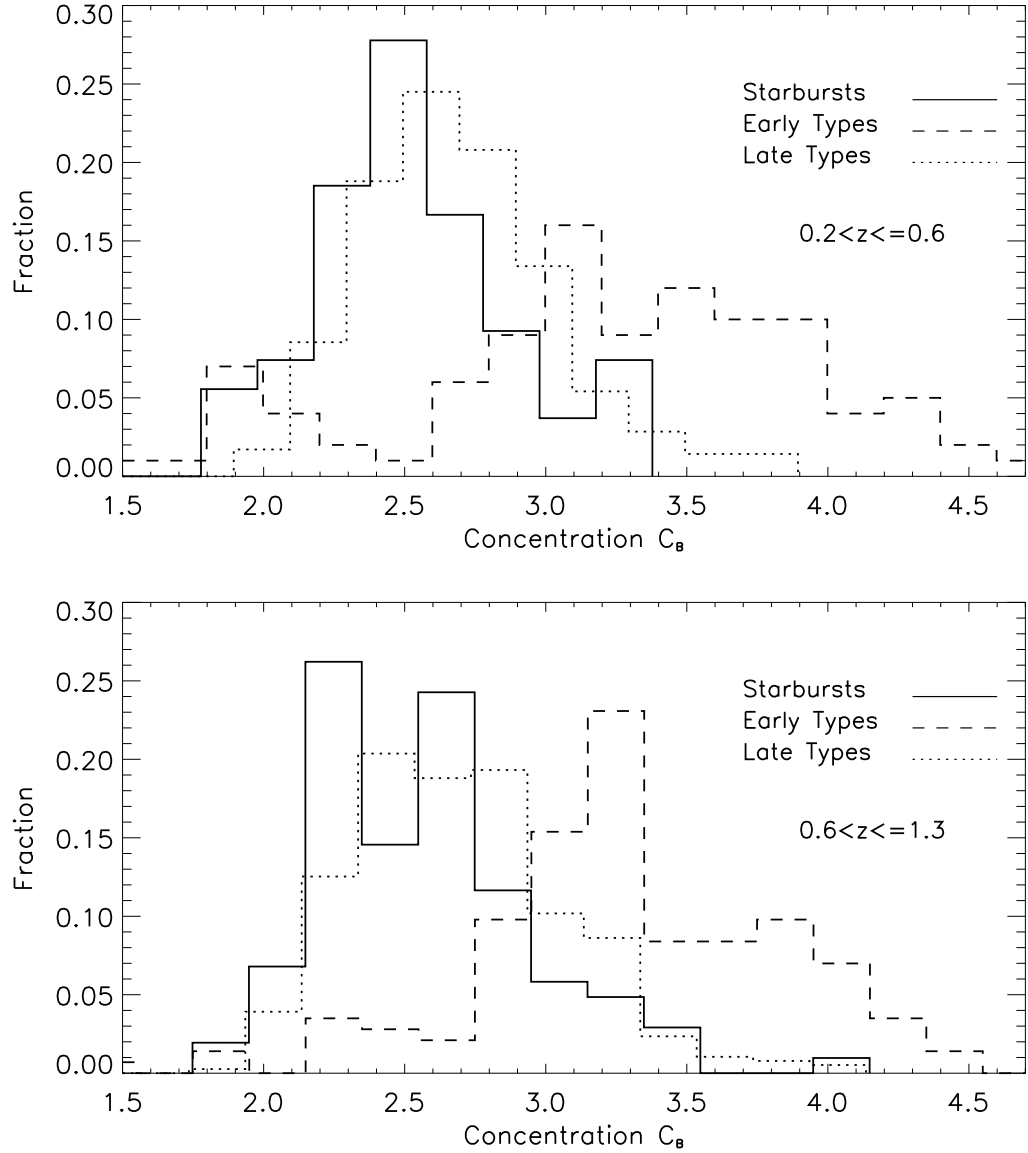,height=6.0in} 
}
\vspace{3pt}
\caption{
The distribution of   concentration indices  ($C_B$) in rest-frame $B$  
for the starburst, late, and early-type galaxies over  two redshift 
intervals which are sampled as outlined in Table 1.
\label{fig1}} 
\end{figure}

\begin{figure} 
\centerline{
\psfig{figure=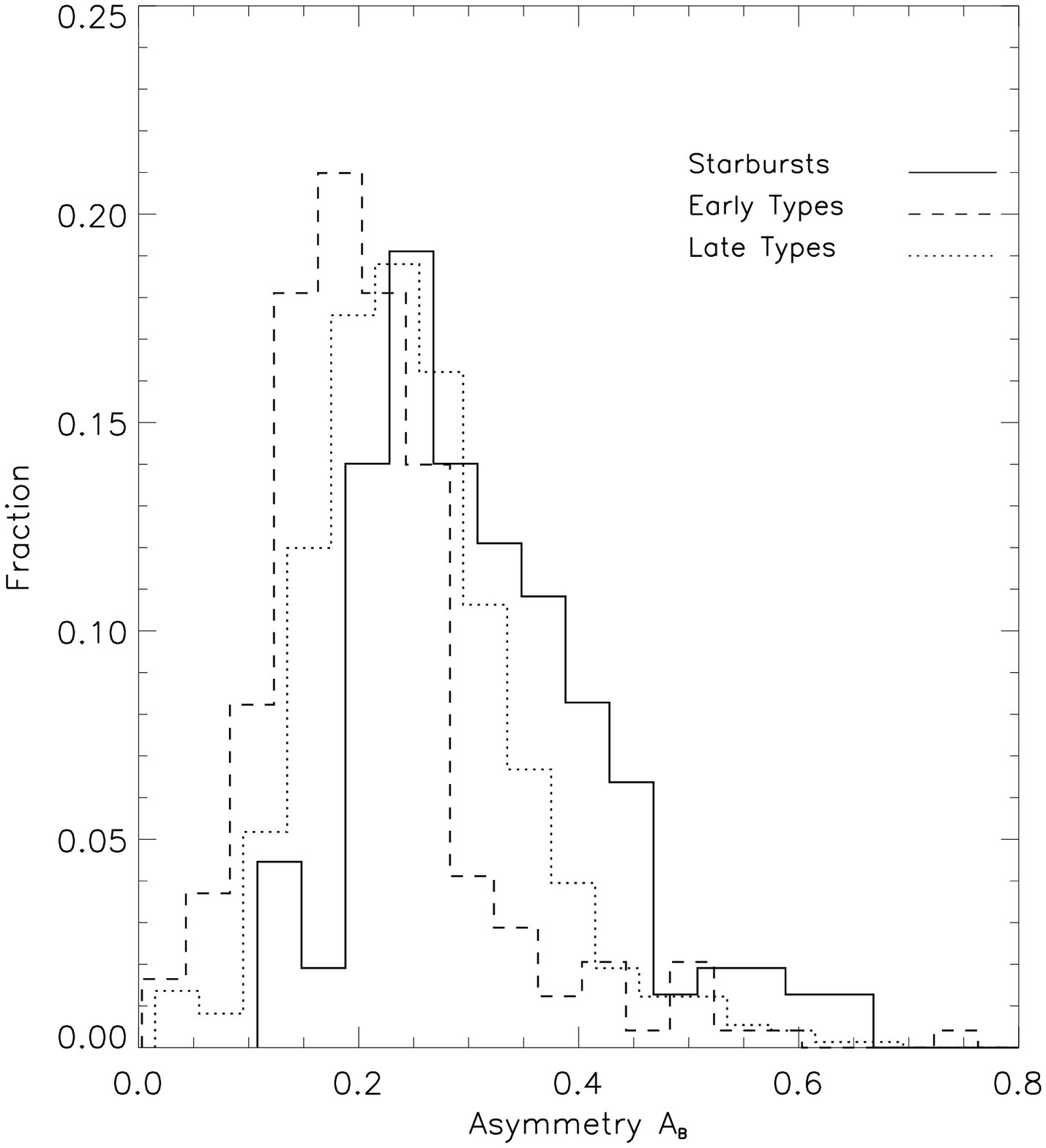,width=3.6in}
\psfig{figure=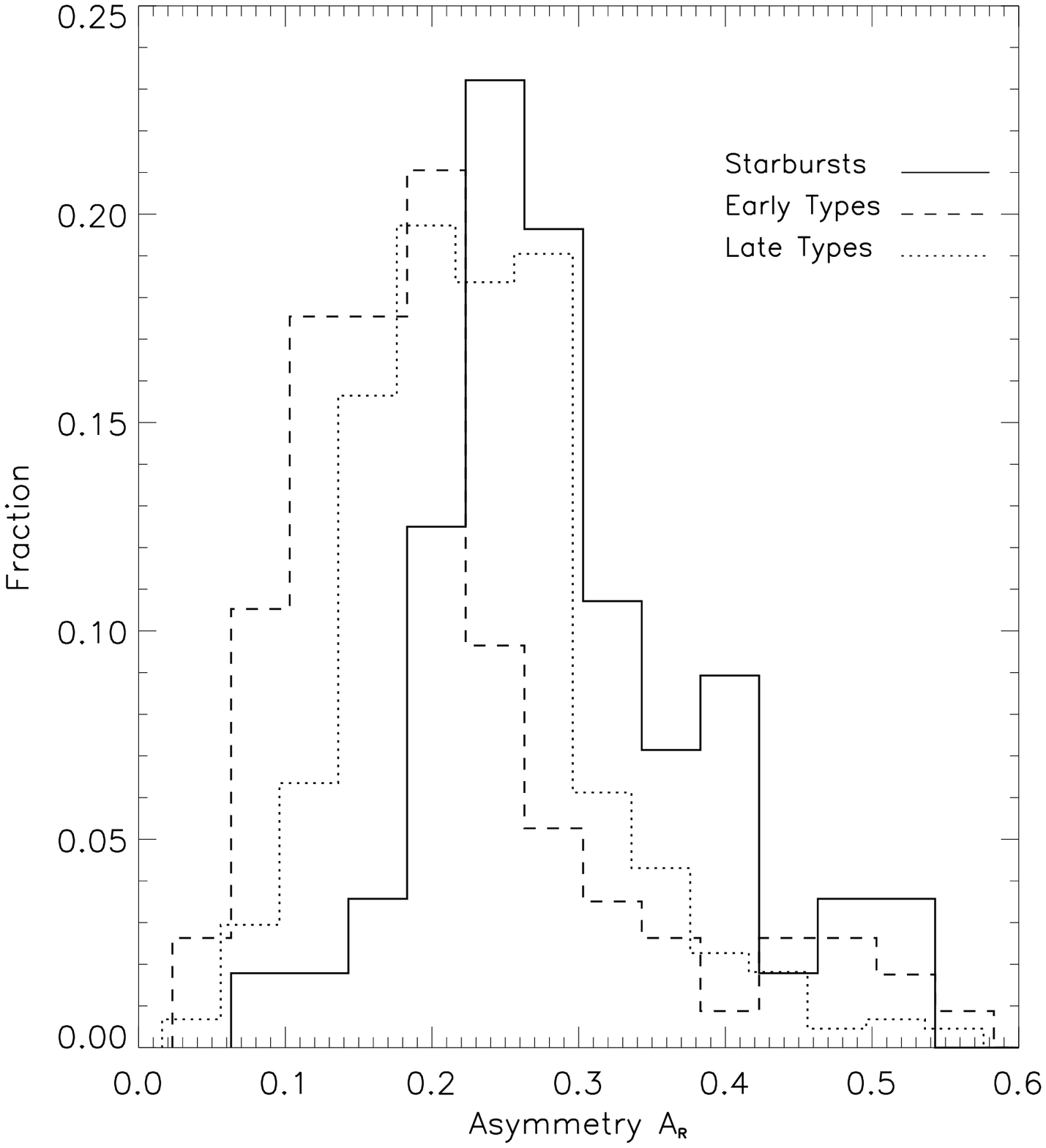,height=3.9in} 
}
\vspace{3pt}
\caption{\it Left:  (a) \rm The distribution of   asymmetry indices 
($A_B$) in rest-frame $B$  for the starburst and control sample.
\it Right: (b) \rm As in (a), but for the rest-frame $R$-band asymmetry indices. The redshift ranges covered  (a) and (b) are listed in Table 1.
\label{fig}} 
\end{figure}

\end{document}